# 'Favorit': **fa**rmers' **vo**latility **ri**sk **t**reatment

Dadasaheb G. Godase[1], P. R. Sheshagiri Rao[2], Anil Gore[3]

1. Shivaji University, Kolhapur
2. President, Chennakeshava Trust, Karnataka; Director, Hive Organics Pvt. Ltd. Bangalore
3. Professor (retired), Department of Statistics, Pune University (SPPU), Pune

1. dggodase7@gmail.com, 2. girickp@gmail.com, and 3. goreanil@gmail.com

**Abstract**

This paper seeks to develop a strategy based on analytics, for an individual Indian farmer to tackle market price fluctuations. The idea is to select a month (or a week or a day) on which to take the produce to market for a good return on the sale. The choice is based on the history of price data and associated variability. Market-wise price data for the last decade or so are used. These ideas are applied to three vegetable crops, tomato, onion, and coriander for some markets in the state of Maharashtra. It is proposed that similar work should be done crop-wise and market-wise for different parts of India by local academic groups from any of the subjects such as statistics, analytics, data science, agriculture, business management, commerce, and economics. The objective is to mitigate the adverse impact of price fluctuation on farmers.

**Introduction**

Farming in India is a risky business. Risk is of many types; rainfall (excess, deficit, bad distribution), pests and diseases, price variation, etc.

Irrigation facilities help fight deficit rainfall. Plant breeders work to develop crop varieties that can withstand pests and diseases. Perhaps the only approach in India to protect farmers from price collapse is the MSP or minimum support price scheme of the central government. Under this scheme, farmers' produce (a particular crop, say wheat) is procured by the government at a specified floor price. This scheme has mainly helped wheat and rice producers. Some state governments may offer support for other crops selectively. As an example, Maharashtra State Government has offered price support for cotton for many years. This year, the open market price of cotton is higher than the support price. Naturally, state government procurement centers are deserted as farmers sell cotton directly to traders at a handsome price.

There is very little protection available for many other crops. One glaring example is tomato. The wholesale price of tomato falls precipitously say below Rs. 400 per quintal, again and again. Newspapers are replete with stories of farmers' dumping truckloads of tomato on the roads in protest (see as an example, news on 27 August 2021 at this link[1]: https://www.livemint.com/news/india/maharashtra-farmers-dump-tomatoes-on-roads-as-prices-crashed-up-to-rs-3-kg-watch-video-11630039021287.html). Similar situation arises with many other crops too. We will consider two more such crops: onion and coriander. The distress and desperation of farmers faced with price collapse make all sensitive citizens concerned and sad.

Price fluctuation is not a problem in India alone. Many studies elsewhere have tried to assess the impact of price volatility on demand and supply of farm produce. Zhang et al.[2] stated that accurate price prediction of agricultural products is useful for planning agricultural production and for developing a balance between supply and demand. Paul Jr et al.[3] studied the effects of price and non-price factors on the supply response of tomatoes in Cameroon. Dragan et al.[4] analyzed changes and future tendencies of price parameters of tomato with descriptive statistics and found that the ARIMA was suitable for price forecasting.

Any effort to mitigate the impact of price crash can surely give relief to farmers. It is worth noting that there is an asymmetry in human reaction to profit and loss. Psychologists have noted[5] that the pain of a loss of an amount is more serious than the pleasure of a gain of a similar amount.

Therefore, the aim of this paper is not to maximize profit but reducing losses. The attempt is to use analytics, not to discover any new macroeconomic insight, but to suggest a small change in a farmer's approach. The target of the study is a farmer who must take his produce to the nearest market as soon as harvest operations are over. Hence, any advisory must be market-specific. Naturally, the present exercise is applicable to a specific crop and a specific market from which price data are sourced. Details must be worked out separately for each market and each crop all over the country. That is why it needs a concerted all-India effort. What we present here is a template for such a coordinated effort.

**Agricultural Marketing in India**

India has a system of organized markets regulated by Agricultural Produce Market Committees to facilitate wholesale trade in farm produce with the stated objective of ensuring a transparent and fair practice to help farmers  There are thousands of such regulated markets (see[6] 
https://dmi.gov.in/Documents/Brief%20History%20of%20Marketing%20Regulation.pdf). 
Information on arrival (in tons), maximum, minimum, and modal price are available on[7] https://agmarknet.gov.in/PriceAndArrivals/DatewiseCommodityReportpart2.aspx.   Variation between markets is quite large. In the case of tomato, in Maharashtra, some markets may have a tiny volume of arrivals (say a couple of tons in a market like Ahmednagar) to a sizable volume (say a few hundred tons in a market like Dindori (Vani)). Chittoor in Andhra Pradesh, perhaps the largest market for tomato in India, can have arrivals as high as 1500 tons in a day. Prices also vary a lot between markets.

**Tomato Price Volatility**

Price data are available on the following link[8]:
http://agmarknet.gov.in/PriceTrends/SA_Pri_Month.aspx

We begin by viewing monthly average prices for Maharashtra. Statewide averages are also available on the website. Our interest is identifying months with maximum average price. We can then contemplate planning the crop so that we approach the market in that month.

| Table 1: Monthly average price (Rs. Per quintal) of tomato in Maharashtra state, by year | | | | | |
|---|---|---|---|---|---|
| Sr. No. | Year | Minimum price | Month | Maximum price | Month |
| 1 | 2011 | 299.26 | August | 1490.33 | January |
| 2 | 2012 | 385.47 | August | 1232.34 | April |
| 3 | 2013 | 610.68 | October | 2237.38 | November |
| 4 | 2014 | 566.35 | May | 2416.82 | July |
| 5 | 2015 | 839.91 | May | 2447.64 | November |
| 6 | 2016 | 315.76 | December | 3917.68 | June |
| 7 | 2017 | 485.05 | January | 4248.34 | July |
| 8 | 2018 | 504.45 | March | 1470.79 | July |
| 9 | 2019 | 795.94 | December | 2281.42 | October |
| 10 | 2020 | 584.56 | May | 2479.07 | September |
| 11 | 2021 | 401.51 | September | 1143.59 | April |

The above table (Table 1) compares monthly average prices within a year and identifies the maximum and minimum price together with the month. Out of 11 years, 3 years get the lowest price in May, 2 years in August and December, and 1 year in January, March, and September. A farmer is likely to face a very low price in the market if he takes his produce to the market in these months. Months absent in this list are February, April, June, July, and November. They seem to be safer months for farmers.

Turning to the maximum price in a year, July gets the maximum price in 3 years, April and November get the maximum price in 2 years each. January, June, October, and September get the maximum price in one year each. The names that do not appear in the list are February, March, May, August, and December. Perhaps farmers should avoid going to market in these 5 months. The month of February is common to the two lists. It does not give a very low price but also does not give a very high price.

In 2011, the lowest price was just under Rs. 300 per quintal while the highest price was almost Rs. 1500 per quintal. A five-fold change. This is bad enough. Check the year 2016. Here the highest price is more than ten-fold of the lowest price! This gives us some idea of the massive fluctuation.

This kind of price crash is not encountered in almost any other business. It is like a Maruti Alto car costing about Rs. 400,000 in one year can be sold only at Rs. 40,000 in another year. It can wipe out even a healthy company.

Since our interest is market-wise analysis, we can view maxima and minima at individual markets instead of those over a region.

A naïve response to this variation in price is to suggest that a farmer should select the month with the best average price and adjust his sowing plans so that the harvest will match with that month.

There are three serious issues with such a strategy.

a. If every farmer (or most of the farmers) selling produce in a given market uses this strategy, there will be a glut in the market in that month and very likely a crash in price.
b. Average price in a month (or week or some such time period) over years, is only part of the story. Variation is the other part. To assess variability, common tools are box plot and confidence interval. They must be used and recommendations adjusted.
c. Though July is an attractive month because of the high average price, shifting to July may not be easy. To get output in July a farmer must sow about 10 weeks earlier which comes out to be April. This is late summer and very few farmers would have access to water to sustain a crop planted in this period. (Such farmers with access to irrigation facilities should indeed consider the shift). Second problem with July is the possibility of heavy rains in late June and early July. Such rains will cause waterlogging, may attract diseases, and decrease yields, particularly in high clay content soils. Perhaps shift to a period closer to one's usual current month of Transplanting may be easier. So, it may be better to indicate how good or bad a month is from a marketing viewpoint and to let a farmer judge if he can move one step to a more attractive month. This approach would require a good understanding of price volatility. Knowledge of maximum and minimum is not enough. A confidence interval would be more useful.

**Confidence Intervals for monthly average prices:** Let us consider a typical market, say, Pimpalgaon in Nashik district in Maharashtra state. We have the average monthly price for tomato available from the website given. We use data since 2011. For each month we have about 11 values. Using these and employing bootstrap sampling, we can generate the distribution of monthly prices and from that derive an empirical confidence interval. We can draw a sample of size 11 from this lot with replacement and get an average price. A whole universe of such average prices can be generated for that month. These values are then arranged in increasing order. To get a 95% confidence interval, we need an upper limit and a lower limit. The upper limit is the average price such that only 2.5% values exceed it. The lower limit is the average price such that only 2.5% values fall below it.

**Fig 1. Confidence intervals for monthly average price (Rs/quintal) for tomato in Pimpalgaon**

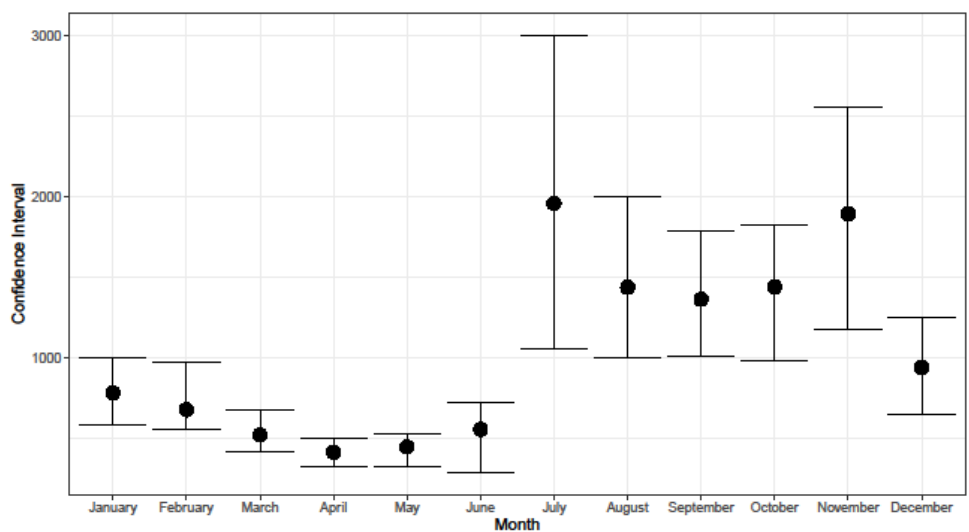

How can farmers use this graph in decision making? Many features of this graph are worth noting. Prices seem to fall into two groups of months. January to June is one group and July to December is another group. First group (say L) shows lower prices but also lower variability. The second group (say H) shows higher means and higher variability. How can we decide about the trade-off between average and variability? In some cases that is easy. Note that even the worst price of July seems to be better than the best price of June. So, it seems safer to plan on going to market in July rather than June.

The question of interest is the generality of these patterns. Can we say the same thing for any market? The answer is negative. Consider the following graph giving confidence intervals for a monthly average price in Satara market. Here also we see two groups. But the first group with low prices is for four months January to April. The prices in May are good. Perhaps December can be included in the first group. The point is that conclusions vary from one market to another. So, the analysis also has to be market-wise. General advice based on the graphs seems to be as follows: If you take your harvest to market in a month in group L, try to move to a month in group H.

To be more specific will need a more precise ranking among months. Our aim is to compare different months and rank them in terms of price.

**Fig 2. Confidence intervals for monthly average price (Rs/quintal) for tomato in Satara**

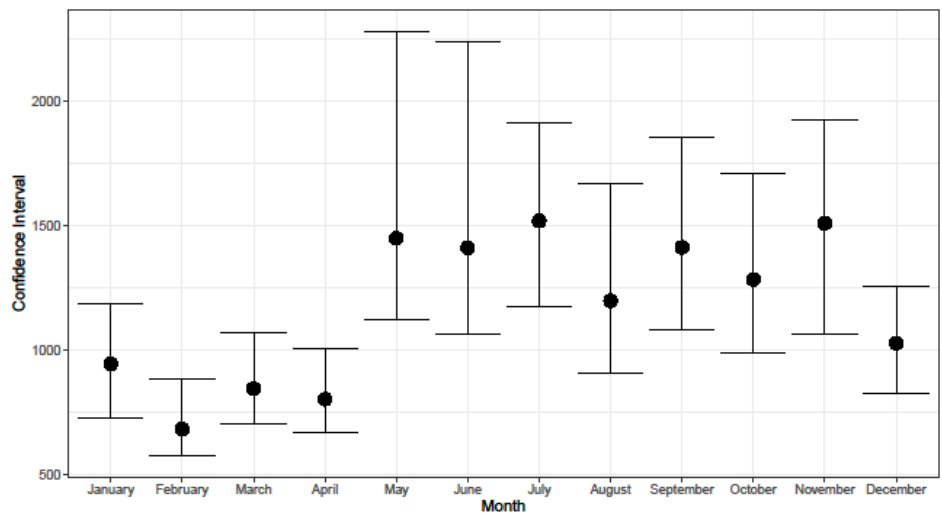

**Ranking Months for Tomato**

Can we rank months of a year on the gradient of sale price for a farmer? As argued earlier, ranking must reflect average price together with volatility. For this purpose, we can use a confidence interval for average price in a month. Let us refer to the confidence intervals calculated for Satara market. The graph is given above. But now we need the actual numerical values of the limits. These are given in the following table (Table 2).

| Table 2: Confidence intervals for monthly average price (Satara market) for tomato (Rs/quintal) based on data 2011-2021 | | | | |
|---|---|---|---|---|
| Month (in alphabetical sequence) | Mean | Lower limit | Upper limit | FLAP Index |
| April | 803 | 659 | 1037 | 2.48 |
| August | 1198 | 861 | 1745 | 1.67 |
| December | 1029 | 764 | 1282 | 2.39 |
| February | 683 | 562 | 898 | 2.52 |
| January | 945 | 746 | 1233 | 2.11 |
| July | 1519 | 1128 | 1944 | 2.12 |
| June | 1411 | 1045 | 2245 | 1.49 |
| March | 845 | 680 | 1081 | 2.40 |
| May | 1449 | 1082 | 2212 | 1.55 |
| November | 1509 | 1067 | 2019 | 1.89 |
| October | 1284 | 978 | 1755 | 1.91 |
| September | 1412 | 1050 | 1923 | 1.93 |

Let us try to rank all 12 months in terms of price. Since ranking a two-dimensional entity is not straightforward, we need to identify some intuitive rules that a ranking procedure should satisfy.

Rule 1: If lower limit of confidence interval for a month A is higher than the upper limit of the confidence interval for month B, then A should have a better rank than B. [verify from the above table (Table 2) that July beats February] This will resolve the question only in some cases. When it doesn't, we need the next rule which is slightly less stringent.

Rule 2: If the mean for month A is higher than the upper limit of the confidence interval for B, then the rank of A should be higher than the rank of B. [verify from the above graph that July beats January]

These two rules may not sort out all comparisons. For resolving the remaining comparisons, we propose an index of the attractiveness of a month.

**Flap Index**: We are trying to combine two dimensions of price (mean and volatility). Let us define a new index to be named FLAP index. FLAP is the acronym of Fluctuation Adjusted Price. We define it as mean/sd. Note that it is the inverse of the Coefficient of Variation (sd /mean). Using this index, we give the last rule.

Rule 3: Month A is ranked higher than month B if the FLAP index of A is higher than that of B.

These three rules together will rank months for a market. A month with a higher rank is more desirable.

Decision recommendation: There may be various constraints on a farmer when it comes to making changes in his cropping rhythm. A crop taken in rabbi season may not be easy to shift to kharif season. So, our approach is to suggest only a small change.

Advice- if the current practice leads to tomato crop coming to market in month A, the farmer should consider months with ranks higher than A, and shift activity so that his harvest comes to the market in one of those months.

| Table 3: Ranking of months in the terms of sale price of tomato for multiple markets in Maharashtra | | | | | |
|---|---|---|---|---|---|
| Rank | Karad | Patan | Satara | Wai | Pimpalgaon |
| 1 | July | July | July | November | November |
| 2 | November | November | May | July | August |
| 3 | October | September | September | October | July |
| 4 | September | May | November | September | September |
| 5 | May | June | June | December | October |
| 6 | June | October | October | January | January |
| 7 | August | January | August | August | December |
| 8 | December | August | December | June | February |
| 9 | January | February | January | May | March |
| 10 | February | December | February | April | June |
| 11 | April | March | April | March | April |
| 12 | March | April | March | February | May |

Note once again that answers vary with the market (see Table 3) and hence the proposed template has to be applied to individual markets to develop recommendations for farmers.

**Problem for Onion**

This is an important vegetable crop for India. Maharashtra (20%), Madhya Pradesh (14%), and Karnataka (12%) are the major producers. Fluctuations in onion prices have caused political upheavals in Maharashtra. Sharad Joshi, a leader of onion farmers in the state, organized their movement demanding fair price[9] (see[10], https://www.news18.com/news/opinion/sharad-joshi-was-first-to-fight-for-economic-freedom-of-farmers-he-would-be-upset-today-2908367.html). In the subsequent election, candidates supported by this movement won many seats in the legislature. It rattled the government.

In 1980, after a fall in onion prices due to an export ban, Sharad Joshi, burst onto the political scene by mobilizing farmers to block the busy Pune-Nashik highway at Chakan in protest. The agitation, which continued for days, and also saw villagers hold up traffic with their bullock carts, led to tremors in Delhi. The government declared that the National Agricultural Cooperative Marketing Federation of India (NAFED) would purchase onions.

Earlier, in March 1978, Joshi had similarly mobilized farmers against depressed onion prices due to an export ban. This agitation led to the Shetkari Sanghatana being launched on 8 August 1979.

Some differences between tomato and onion as crops need to be noted. a) Onion is planted in Rainfed lands to a large extent in Khrif season which makes it very vulnerable to unexpected wet and dry spells; b) It has a much higher shelf life and is easier to transport; c) Large volumes are imported and exported depending on government policy.

Lasalgaon in Nashik District (Maharashtra) is the largest market for onion in the country (see[11], https://en.wikipedia.org/wiki/Lasalgaon). Onion is different from tomato in that it has only limited sowing windows. Hence, the options available to a farmer are limited. He must choose one of these. We have studied two sowing seasons for the onion crop in India. They are Kharif (planted between June-July and harvested in September-October); Rabi (planted between October-November and harvested in February-March. (Details vary from one region to another. In some cases, there are three sowing seasons. The approach developed here continues to apply there too.). Note that it is possible to rank months by prices of onion using an approach developed above for tomato. However, while it may be useful for traders, it may not be relevant for farmers since, unlike tomato, they cannot choose the sowing season at will.

Let us assume that a farmer sticks to one of the above two seasons for sowing. Then, the option available to him is among weeks in the window of choice. Further, in kharif season, if the decision to sow is based on rains, then the choice may be restricted even more. Hence, as a template, we can only consider the relative merit of sowing in different weeks in each season. We will check prices and rank weeks in September-October period and February-March period.

We need not repeat the details. It should suffice to say that the data for Lasalgaon market in Nashik district of Maharashtra state was used. Results are given in two tables (Table 4 and Table 5) for two seasons.

It appears that for the kharif season, it is better to be early and plant in September. On the other hand, for rabbi season, it seems more profitable to be late and plant in March.

| Table 4: Ranking of weeks for kharif onion, Lasalgaon market, Maharashtra | | |
|---|---|---|
| Week | Rank | Month |
| 1 | 7 | September |
| 2 | 8 | September |
| 3 | 6 | September |
| 4 | 5 | September |
| 5 | 2 | October |
| 6 | 1 | October |
| 7 | 3 | October |
| 8 | 4 | October |

| Table 5: Ranking of weeks for rabbi onion, Lasalgaon market, Maharashtra | | |
|---|---|---|
| Week | Rank | Month |
| 1 | 1 | February |
| 2 | 2 | February |
| 3 | 3 | February |
| 4 | 5 | February |
| 5 | 4 | March |
| 6 | 8 | March |
| 7 | 6 | March |
| 8 | 7 | March |

**Problem for Coriander**

The final case we take up is that of coriander. This is a favorite condiment among Indian chefs. Coriander takes about 3 months from sowing to harvest. We will only consider farming for leaves. Coriander seeds are also a product, but the dynamics involved may be different. For marketing fresh green leaves, there is a short window of say 1 or 2 weeks once the crop is ready. Later, there is growth of flowers making the product less desirable to consumers. So, the decision to be made is selection of a day out of (say) the 8 days to take the crop to market, once the crop is ready.

Again, it is possible to rank months for coriander. Instead, we focus on the available window of 8 days to choose from. This short decision window is peculiar to coriander. In the example chosen, the price variation may not seem large. But if one can anticipate a good day in the market, it signifies additional income at virtually no additional burden. So, here our attempt is to anticipate daily prices in the market of interest and based on them, recommend a day on which to take coriander crop to the market.

We fit a time series model (ARIMA) to daily prices of the period of 100 days. This model is used to forecast prices for the next 8 days. Advice emerging from the modeling exercise is that a farmer should select a day with the highest forecast price to take his produce to market.

A key question is whether this approach will work. The problem of predicting vegetable prices continues to engage scientists all over the world. Helin et al.[12] in Korea tried to predict monthly prices and found that high volatility is a tough problem. In 2018, Kaggle the well-known portal that runs analytics competitions asked the question 'what is the best model to *predict* the *prices* of agriculture produce?' (see[13], https://www.kaggle.com/datasets/raghu07/vegetable-and-fruits-price-in-india). Recently, Purohit et al.[14] forecasted prices of TOP (tomato, onion, and potato) in India. They tried out a whole range of approaches and suggested that time series models often produce poor forecasts.

So, the literature warns users that ARIMA models may not do a good job in predicting vegetable prices. In the light of such signals, it is necessary to verify that any method proposed indeed does a reasonably good job. The usual approach to this problem is to compare predicted and actual prices. We have taken two data sets from two markets and given predicted as well as true values.

**PRIM Criterion for Performance of a Time Series Forecasting Model**

Our contention is that matching between the two sets of prices would be useful but not necessary for our purpose. The aim is not to guess future prices correctly but to give the farmer a date on which he should take his produce to market. A farmer who follows this advice gets a certain price. What is it to be compared with? Naturally, the comparison should be with the price that was yielded by the traditional practice of farmers. What is the traditional practice? The farmer chooses a date randomly among the available dates. In the case of coriander, the available option is a window of 8 days after the crop reaches maturity in the field. Hence, the expected realization would be the arithmetic average of prices for those eight days. This is the benchmark. Any alternative course of action should improve upon this value. Our recommendation is to go to market on the day for which the predicted price is the highest among the predictions for eight days. So, the realization achieved using this recommendation is the actual price of that day. The forecasting approach should be considered useful to the farmer if this actual price is higher than the average true price for eight days. This is the proposed PRIM (price realization improvement) criterion. Let us apply this criterion to the case of forecasting the price of coriander. We have considered two illustrative markets, Solapur, and Kolhapur. In each case, we have generated predictions based on ARIMA. Dates, predicted and actual prices are all given in the table below (Table 6).

| Table 6: predicted and actual prices of coriander in two markets in Maharashtra |||||||
| Solapur Market ||| Kolhapur Market |||
| Date | Predicted Price (Rs./quintal) | Actual Price (Rs./quintal) | Date | Predicted Price (Rs./quintal) | Actual Price (Rs./quintal) |
| --- | --- | --- | --- | --- | --- |
| 28-06-2021 | 538.85 | 700 | 17-09-2021 | 4748.04 | 5250 |
| 29-06-2021 | 443.97 | 500 | 18-09-2021 | 4918.73 | 7000 |
| 30-06-2021 | 477.58 | 400 | 19-09-2021 | 4152.19 | 3500 |
| 01-07-2021 | 482.97 | 400 | 20-09-2021 | 4375.35 | 5250 |
| 02-07-2021 | 469.54 | 300 | 21-09-2021 | 4700.03 | 6300 |
| 03-07-2021 | 471.68 | 400 | 22-09-2021 | 5117.38 | 7700 |
| 04-07-2021 | 422.24 | 452 | 23-09-2021 | 4031.60 | 5950 |
| 05-07-2021 | 451.98 | 400 | 24-09-2021 | 3187.78 | 5250 |
|  | Mean | 444 |  | Mean | 5775 |

In the case of the Solapur market, the highest predicted price is on 28-06-2021. For our forecast to be useful the average true price should be lower than the true price on that date. Note that the average true price is Rs. 444 while the true price on the recommended day is Rs, 700. So, a farmer who follows the advice based on the prediction stands to gain handsomely. In the second example namely the Kolhapur market, the highest predicted price is on 22-09-2021. The true price on that day is 7700 while the average true price for the eight days is 5775. Again, our guidance would be beneficial to the farmer.

**Adoption of the Method by Farmers**

If the proposed approach to selecting the day for taking coriander (or any other vegetable) to market is to be put into practice, some operational mechanism would have to be put in place. We see two such possibilities. One is to collaborate with a newspaper or any other media which agrees to publish relevant predications to be generated daily/weekly or at suitable time intervals. For this to be done for each market would be a massive task and an all-India network of data analysts would have to share the work. The other possibility is to develop an application suitable for a cell phone. Any interested farmer should be trained to use the app to generate predictions and use them to choose the day on which to take his produce to market. This approach also involves a substantial amount of work. Our plan is to see if the method is found attractive by farmers and then commit ourselves to the next stage of action.

**Problem of Arbitrage**

Arbitrage is the practice of taking advantage of a difference in prices in two or more markets. In an efficient economic system, prices across markets should be similar. If one market has a higher price, sellers, suppliers, producers should shift part of the product to the market with a higher price thus bringing about approximate equality. It appears, however, that there may be anomalous differences in price across adjacent markets, not explained by distance/transport cost.

Consider three adjacent markets Satara, Patan, and Karad. Distances between them are less than 100 km; Transport cost per kg of tomato is about Rs. 4. So, if the month with Rank 2 is November in Satara and May for the other two markets that sounds anomalous. As market prices are well known, will not the traders simply move the produce to the market at higher price? This issue needs a detailed investigation and is being attempted.

Is this because of the lack of consistency of price differences? If the price difference evaporates in a day or two, no one will want to risk shifting to a different market. Distance, transport cost, and persistence of price difference will determine if it is a real arbitrage opportunity. Attractiveness will also depend on the quantum of difference. Then there is the long-distance operation. Tomatoes in Maharashtra are bought locally and transported all the way to destinations in Haryana and Delhi. Apples from Himachal Pradesh are transported all over south India. Are price differences in such cases similar or different from the common trading very close to the locality of production? All such matters need a separate investigation.

**Discussion**

The aim of this paper is to offer advice to an individual farmer taking his produce to the nearest market, about possible action to avoid low prices. We advise on the best bet Sowing/Transplanting window. Such advice is available well ahead of action to be taken; providing sufficient lead time for farmers to make their choice and implement the decision. Three different crops have been considered. In the case of tomato, a method is offered to rank months in terms of price prospects. In the case of onion, weeks during which the crop has to go to market are ranked. In the last case of coriander, advice is about the price to expect in the next 8 days. Clearly, the decision support templates depend on the crop of interest, specific decisions required, and dynamics of the relevant markets. So, such analysis and advice have to be customized to both specific crops as well as their different growing regions. The approach is rather simplistic in the sense, only one variable (wholesale price) is taken into consideration.

Other variables obviously play the part. How critical they are depends on the specific context, crop, and region. As an example, productivity changes with the season. Lower price may end up giving higher profits if higher productivity more than compensates for lower price. Second point is that a very narrow view of the neighborhood market may not always be appropriate. There is always an arbitrage opportunity to go to a distant market that offers a higher price if the price difference can cover transport cost. For Kolar-Tumkur area of Karnataka, for tomato, February and March planting is the best bet for an assured high price. Even though yield per acre decreases due to high temperature, the higher price compensates for it. High summer temperature prevalent in the other major Tomato growing regions of Maharashtra, Andhra Pradesh etc. constrains Tomato cultivation there. Hence, in June and July, there is high demand but short supply -an ideal opportunity for Kolar-Tumkur region. They have successfully used this opportunity for over a decade.

Then there is the aspect of weather prediction. Climate knowledge can be used to anticipate price movement. Events such as El-Nino can be predicted 2-3 months ahead of the planting season. These events have a high correlation with droughts and flooding (Gadgil et al.[15]). In Anantpur region of Andhra Pradesh with a seasonal average rain of 42 cm, the chance of high rainfall is 62% in El-Nino+1 years, while it is a mere 9% during El-Nino years. In years of High rainfall (>50 cm), long and intense wet spells occur September, October months leading to waterlogging and soil borne diseases in all three crops. In El-Nino years, the chance of low rainfall (<30cm) is 30%; while it is a mere 14% during El-Nino+1 years. Long dry spells occur in September and October months leading to moisture stress in rainfed farms promoting certain pests.

Like El-Nino, we have to identify other weather events that allow sufficient lead time for the farmers to choose the best options. Our Nation has a unique advantage of long-term daily climate data at fine resolution (1000 sq km grid). As in the case of the market price of crops, context-specific utilization of this wealth of data in active partnership with the users (farmers, traders) will generate many options to enhance productivity, reduce the risk of pests and diseases.

Farm product prices are impacted by a diverse set of factors depending on the crop and region. One illustrative factor is the availability of substitutes to the crop we focus on. In some regions of south India, Tamarind is a substitute for Tomato and is available in plenty during its harvest season of December-February. This substitution dampens the demand for Tomato in this region which will be reflected in the price.

Our efforts and discussion belong to the broad area of anticipatory management of farming. Not only price, but all aspects of farming can benefit from anticipation. Along with the weather, it is possible to predict pest attacks and diseases and manage them economically. Anticipation of good and bad events will help us design resilient farming systems that are context-specific to the agro-eco system. During bad times, inputs can be designed for reducing risks of loss in yields, controlling damage by pests, and diseases. During good times, we can seek high yields through best options in farming systems and inputs. As both weather and market prices, the biggest risk in farming. are beyond the control of farmers – anticipatory management for these critical factors is the best way forward. Such SMART anticipatory management is very likely to lead to lower use of pesticides, fertilizers, irrigation, and to increase biodiversity, thus enhancing sustainability. Scientists and entrepreneurs must join hands with farmers to create a new ecosystem collaborative and context-specific action research that leverages the technology for a win-win-win situation to benefit Farmers, Entrepreneurs, and Environment.

The present paper is a baby step in that direction.

**Acknowledgments**

The authors are thankful to Dr. Niranjan Joshi and Dr. Anand Karandikar for their critical comments and suggestions.